\documentclass[%
 aip,
 amsmath,amssymb,
 reprint,%
groupedaddress
]{revtex4-1}

\usepackage{graphicx}% Include figure files
\usepackage{dcolumn}% Align table columns on decimal point
\usepackage{bm}% bold math
\usepackage[T1]{fontenc}
\usepackage[english]{babel}
\usepackage{mathptmx}
\usepackage{xcolor}

\newcommand{\dd}{\mathrm{d}}

\newcommand{\be}{\begin{equation}}
\newcommand{\ee}{\end{equation}}
\newcommand{\bea}{\begin{eqnarray}}
\newcommand{\eea}{\end{eqnarray}}

\newcommand{\revision}[1]{\textcolor{black}{#1}}

\begin{document}

%  \preprint{AIP/123-QED}

\title[]{Overtwisting induces polygonal shapes in bent DNA}
% Force line breaks with \\

\author{Michele Caraglio}
\author{Enrico Skoruppa}
\author{Enrico Carlon}
\email{enrico.carlon@kuleuven.be}
\affiliation{Laboratory for Soft Matter and Biophysics, KU Leuven, Celestijnenlaan 200D, 3001 Leuven, Belgium}

\date{\today}% It is always \today, today,

\begin{abstract}
\revision{
By combining analytical results and simulations of various coarse-grained
models we investigate the minimal energy shape of DNA minicircles which
are torsionally constrained by an imposed over or undertwist. We show that
twist-bend coupling, a cross interaction term discussed in the recent DNA
literature, induces minimal energy shapes with a periodic alternance of
parts with high and low curvature resembling rounded polygons. We briefly
discuss the possible experimental relevance of these findings. We finally
show that the twist and bending energies of minicircles are governed by
renormalized stiffness constants, not the bare ones.  This has important
consequences for the analysis of experiments involving circular DNA
meant to determine DNA elastic constants.}  
\end{abstract}

\maketitle

%%%%%%%%%%%%%%%%%%%%%%%%%%%%%%%%%%%%%%%%%%%%%%%%%%%%%%%%%%%%%%%
%%%%%%%%%%%%%%%%%%%%%%%%%%%%%%%%%%%%%%%%%%%%%%%%%%%%%%%%%%%%%%%
%%%%%%%%%%%%%%%%%%%%%%%%%%%%%%%%%%%%%%%%%%%%%%%%%%%%%%%%%%%%%%%
\section{Introduction}

At mesoscopic length scales the elastic response of double stranded
DNA to mechanical stresses is usually described by the twistable
worm like chain (TWLC), which is characterized by just two elastic
stiffnesses corresponding to bend and twist deformations respectively
\cite{mark15,moro97}. At length scales of several base pairs,
relevant to protein-DNA interactions, the sequence dependent geometry
of the double helix leads to a rich spectrum of elastic properties
\cite{lank03,lave10,pasi14}. Even at those length scales, however, it
is useful to employ simplified, analytically treatable representations
that allow for the identification of generic features that are otherwise
masked by sequence dependent variations.  A suitable model that takes into
account the general features of DNA geometry while still fulfilling the
requirement of analytical treatability was put forward in the mid-90's
by Marko and Siggia~\cite{mark94}, who derived it from the analysis of
the molecular structure of DNA.  These authors showed that the asymmetry
between major and minor grooves generates a coupling between bend and
twist~\cite{mark94}, supplementing the TWLC with an additional stiffness
parameter.  Symmetry implies the existence of twist-bend coupling, but
it does not yield any information about the magnitude of the associated
coupling constant.  For a long time, the effect of twist-bend coupling
has been ignored, assuming that such coupling would have a minor influence
on DNA elastic properties.  However, a recent comparative analysis of two
very similar coarse-grained DNA models one with symmetric grooves and one
with asymmetric groove has shown that the twist-bend coupling constant
is comparable in magnitude to the other elastic parameters describing
bending and torsional stiffness~\cite{skor17}. A similar conclusion was
drawn from the analysis of all atom simulations of DNA~\cite{lank03}.

Twist-bend coupling has been shown to influence elastic properties
of DNA both at long~\cite{nomi17,nomi18} and short~\cite{skor18}
molecules. For instance, in $\approx 100$ base pairs bent DNA fragment
twist-bend coupling induces sinusoidal standing waves in the twist
\cite{skor18}. These waves are comparable in shape and magnitude to
those observed in nucleosomal DNA, which is wrapped around histone
proteins~\cite{rich03}. In longer kilobase pair DNA subject to a
stretching force and to a torque, as in single molecule experiments,
twist-bend coupling has been shown to lead to a rescaling of the elastic
parameters~\cite{nomi18}. 
The aim of this paper is to focus on minimal energy shapes of DNA
minicircles of about $10^2$ base pairs which are overtwisted. We show,
by combining analytical and numerical results, that twist-bend coupling
induces distinctive shapes of DNA in which the curvature is periodically
modulated in an alternance of high and low bending regions. The
periodicity is close to that of a straight double helix, but depends
on the degree of overtwisting. 

%  While in the cell or under typical in-vitro experimental
%  conditions DNA is subject to considerable thermal fluctuations, it is
%  nonetheless interesting to understand ground state shapes of the double
%  helix. 
\revision{
Thermal fluctuations strongly influence the conformation of linear DNA
molecules of lengths exceeding the bending persistence length, which
is approximately $50$~nm. In shorter, constrained and highly bent DNA,
a situation of relevance in DNA-proteins complexes, thermal fluctuations
do not influence strongly the shape of the molecule, which in first
approximation assumes its minimal energy conformation. Elastic energy
minimization has indeed been used to obtain the shape of DNA looping
out of a Lac operon~\cite{bala99} or for DNA wrapped around histone
proteins~\cite{moha05}. Although in this paper we focus on minimal energy
shapes of free standing over and undertwisted minicircles, we expect that
our analysis will be of particular relevance for short constrained DNA,
i.e. partially bound to proteins, where thermal fluctuations have a small
effect. For convenience we focus here on homogeneous minicircles, which
are simpler to simulate and to describe analytically. We expect, however,
that the shapes discussed here also applies to other more complex cases,
such as DNA looping where translational invariance is broken.}

%%%%%%%%%%%%%%%%%%%%%%%%%%%%%%%%%%%%%%%%%%%%%%%%%%%%%%%%%%%%%%%
%%%%%%%%%%%%%%%%%%%%%%%%%%%%%%%%%%%%%%%%%%%%%%%%%%%%%%%%%%%%%%%
%%%%%%%%%%%%%%%%%%%%%%%%%%%%%%%%%%%%%%%%%%%%%%%%%%%%%%%%%%%%%%%
\section{DNA elasticity}

\begin{figure}[t]
\includegraphics[scale=1]{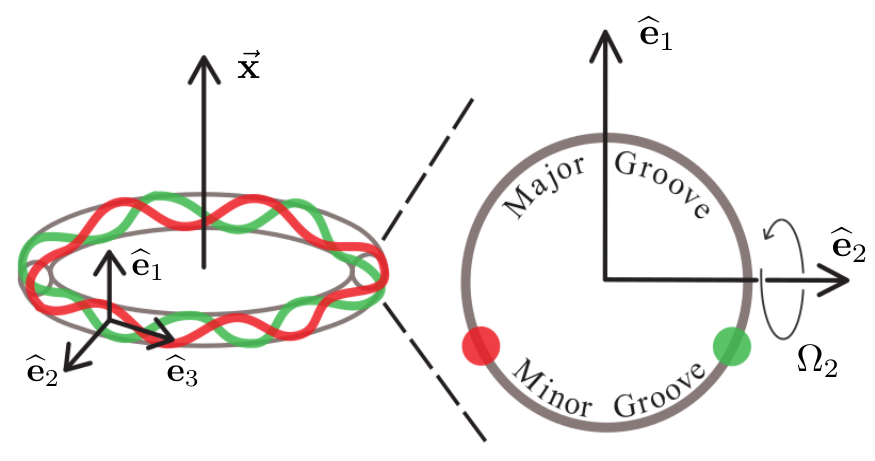} 
\caption{The DNA molecule 
\revision{(sketched as two intertwined strands - red and green)}
is modeled as a continuum
elastic rod in each point of which an orthonormal frame is attached.
${\widehat{\bf e}}_3$ is tangent to the curve while ${\widehat{\bf
e}}_1$ and ${\widehat{\bf e}}_2$ lie on the plane of the ideal planar
Watson-Crick bases. In the right panel, 
\revision{${\widehat{\bf e}}_3$ points into the plane of the paper and
$\Omega_i$ specifies the rotation giving the next triad.
The rotation around ${\widehat{\bf e}}_2$ represented by the curly arrow
corresponds to a positive $\Omega_2$ (i.e. a bend towards the major groove).}
}
\label{FIG:intro} 
\end{figure}

We briefly recall the formalism used to describe a twistable polymer such
as DNA (see also e.g. Ref.~\onlinecite{brac14} for more details) and review some
properties of the model with twist-bend coupling, focusing on the bending
and torsional persistence lengths.  The conformation of an inextensible
twistable elastic rod can be parametrized by three strain fields $\Omega_1
(s)$, $\Omega_2 (s)$ and $\Omega_3 (s)$, where $0 \leq s \leq L$ is a
curvilinear coordinate and $L$ the total length of DNA. $\Omega_1$ and
$\Omega_2$ are bending densities also referred to as ``tilt'' and ``roll''
deformations, while $\Omega_3$ is the excess twist density. DNA has an
intrinsic twist density $\omega_0= 2\pi/l_0 \approx 1.85~\text{nm}^{-1}$,
with $l_0=3.4~\text{nm}$ corresponding to the distance of one helical
turn. Given the $\Omega_i(s)$ the three dimensional shape of the molecule
can be obtained by solving the differential equations $i=1,2,3$:
\begin{equation}
\frac{\dd{\widehat{\bf e}}_i}{\dd s} = 
\left[ \Omega_1 {\widehat{\bf e}}_1 +
\Omega_2 {\widehat{\bf e}}_2 +
\left(\Omega_3+\omega_0 \right)  {\widehat{\bf e}}_3
\right]
\times {\widehat{\bf e}}_i \; ,
\label{eq:rotation}
\end{equation}
where $\{{\widehat{\bf e}}_1(s), {\widehat{\bf e}}_2(s), {\widehat{\bf
e}}_3(s)\}$ defines an orthonormal triad (the Darboux frame) where
${\widehat{\bf e}}_3$ is tangent to the curve, while ${\widehat{\bf
e}}_1$ and ${\widehat{\bf e}}_2$ lie on the plane of the ideal, planar
Watson-Crick base pairs \cite{mark94}, see Fig.~\ref{FIG:intro}. By
convention ${\widehat{\bf e}}_1$ is directed along the symmetry
axis of the two grooves and points in the direction of the major
groove. Finally ${\widehat{\bf e}}_2 = {\widehat{\bf e}}_3 \times
{\widehat{\bf e}}_1$, which yields to a vector connecting the two DNA
backbones~\cite{olso01}. The conformation with all $\Omega_i=0$ describes
a straight twisted rod with an intrinsic twist density $\omega_0$.

%%%%%%%%%%%%%%%%%%%%%%%%%%%%%%%%%%%%%%%%%%%%%%%%%%%%%%%%%%%%%%%%%%%%%%%%
%%%%%%%%%%%%%%%%%%%%%%%%%%%%%%%%%%%%%%%%%%%%%%%%%%%%%%%%%%%%%%%%%%%%%%%%

From the analysis of the molecular symmetry of DNA, Marko and Siggia
\cite{mark94} derived the following free energy functional to lowest
order in $\Omega_i$:
\begin{equation}
\beta E = \frac{1}{2}\int_0^L \left(
A_1 \Omega_1^2 + A_2 \Omega_2^2 + 
C \Omega_3^2 + 
2G \Omega_2 \Omega_3
\right) \dd s \; ,
\label{eq:energy}
\end{equation}
where $\beta=1/k_BT$ is the inverse temperature, and $A_1$, $A_2$, $C$
and $G$ are the stiffness parameters. The last term in~\eqref{eq:energy}
couples twist ($\Omega_3$) with the bending towards the DNA grooves
($\Omega_2$). In the case of vanishing twist-bend coupling constant $G=0$
one recovers the twistable wormlike chain (TWLC) which is the commonly
employed model to describe DNA elasticity. In this paper we focus in
particular on effects of twist-bend coupling on the minimal energy
shapes of bent DNA.  Before entering into the details we recall
in this Section some properties of the model~\eqref{eq:energy}, which
differ from those of the standard TWLC.

The bending persistence length $l_b$ of the model~\eqref{eq:energy} has
been computed from the decay of the tangent-tangent correlation and it
is given by the following relation~\cite{nomi17,skor17}
\begin{equation}
\frac{1}{l_b} = \frac{1}{2} \left( \frac{1}{A_1} + 
\frac{1}{\widetilde{A}_2}\right) \; ,
\label{lb}
\end{equation}
where
\begin{equation}
\widetilde{A}_2 = A_2 \left( 1 - \frac{G^2}{A_2 C}\right)
\label{shiftA2}
\end{equation}
can be viewed as a rescaled bending stiffness along the "easy" bending
axis (recall that $A_2 < A_1$).  In the isotropic TWLC limit $A_1=A_2=A$
and $G=0$ Eq.~\eqref{lb} reduces to the well-know results $l_b=A$, i.e.\
$A$ is both the bending stiffness and the bending persistence length. In
the anisotropic case $A_1 \neq A_2$ and $G=0$, Eq.~\eqref{lb} shows that
$l_b$ is the harmonic mean of $A_1$ and $A_2$, a known result,
see Ref.s~\onlinecite{lank00,esla08}. 
In the general case $G \neq 0$ the result can be
cast in the form of a harmonic mean of $A_1$ and $\widetilde{A}_2$. From
the analysis of the twist correlations one can calculate the twist
correlation length \cite{nomi17} and finds $l_t = 2 \widetilde{C}$ with
\begin{equation}
\widetilde{C} = C \left( 1 - \frac{G^2}{A_2 C}\right) \; ,
\label{lt}
\end{equation}
which reduces to the well-known result $l_t = 2C$ in the TWLC limit $G=0$.
Equations~\eqref{lb} and~\eqref{lt} are exact and have been obtained from
the analysis of thermal fluctuations of the model~\eqref{eq:energy}.
These quantities are also relevant in minimal energy DNA conformations
that are permanently bent and twisted, as we will show in the next
Sections.

\section{DNA minicircles}
\label{sec:minicircles}

A considerable amount of studies have been devoted to the
analysis of equilibrium and kinetic properties of DNA minicircles
\cite{moha03,olso04,du05,moha05,lank06,noro08,demu09,lion11,wang16,schu17,pasi17}
or to short DNA loops obtained by bending two ends of 
DNA~\cite{sank05,mull15}. Our aim is to describe the effect that
twist-bend coupling has on the minimal energy shape of overtwisted
minicircles. We employ a homogeneous model~\eqref{eq:energy}, neglecting
sequence-dependent variations of the elasticity, which we expect to
be valid when averaging over different sequences.  In a minicircle, the
two \revision{ends} of the double helix are covalently sealed and in order for the
loop to close one usually needs to over or underwind the molecule so
that the end point meet in phase.  An excess linking number can also
be induced by using appropriate enzymes that overtwist the molecule. In
what follows we discuss separately the torsionally relaxed and the over
and under-twisted cases.

\subsection{Torsionally relaxed minicircles}

As this case has been discussed recently in Ref.~\onlinecite{skor18}, 
we just briefly review it here. 
An analytical shape for anisotropic bending stiffness and
$G \neq 0$ cannot be found easily. The constraint to be imposed to form
circular DNA requires that the end points of the strands meet each other
smoothly and that the tangent vector is continuous. This constraint is
typically expressed using lab-frame quantities, as Euler angles describing
the configuration of the DNA with respect to some fixed axes. The minimal
energy shapes could then be calculated numerically. Here, however,
we employ an approximation that allows us to perform a local energy
minimization, expressing the constraint in terms of $\Omega$'s. We assume
that the exact solution is a small perturbation of a perfect circle and
that the solution is periodic over helical repeats.

The vector $\vec{\Omega}_b \equiv \Omega_1 {\widehat{\bf e}}_1 + \Omega_2
{\widehat{\bf e}}_2$ is the bending density.  Given a fixed unit vector
${\widehat{\bf x}}$ (see Fig.~\ref{FIG:intro}), Ref.~\onlinecite{skor18}
introduced the following local constraint $-\mu~\vec{\Omega}_b \cdot
{\widehat{\bf x}}$, with $\mu$ a Lagrange multiplier. This term favors
the alignment of $\vec{\Omega}_b$ along ${\widehat{\bf x}}$, i.e. it
forces the molecule to be bent and to remain close to a plane orthogonal
to ${\widehat{\bf x}}$ (see Fig.~\ref{FIG:intro}).  For a straight
double helix in the plane orthogonal to ${\widehat{\bf x}}$ and with
constant twist density equal to $\omega_0$, one can choose the curvilinear
coordinate such that ${\widehat{\bf x}} = \sin( \omega_0 s) {\widehat{\bf
e}}_1+ \cos (\omega_0 s) {\widehat{\bf e}}_2$.  This relation remains
approximately valid if within one helical turn bending is weak and local
twist variations are small, conditions which can be respectively expressed
as $|\vec{\Omega}_b| \ll \omega_0$ and $|\Omega_3| \ll \omega_0$. Within
this approximation the constraint takes the following form:
\begin{eqnarray}\label{Eq:OmegaMinicircle}
\beta \widehat{E} &\equiv& \beta E 
- \mu \int_0^L \vec{\Omega}_b \cdot {\widehat{\bf x}} \, \dd s
\nonumber \\
&=&
\beta E - \mu \int_0^L  \left[
\Omega_1 {\widehat{\bf e}}_1\cdot{\widehat{\bf x}} +
\Omega_2 {\widehat{\bf e}}_2\cdot{\widehat{\bf x}} 
\right] \, \dd s
\nonumber \\
&\approx&
\beta E - \mu \int_0^L \left[ \Omega_1 \sin( \omega_0 s) + 
\Omega_2 \cos (\omega_0 s) \right] \dd s \; ,
\label{lagrange1}
\end{eqnarray}
with $E$ given by~\eqref{eq:energy}. Minimization with respect to
$\Omega_i$ gives the following solution for a torsionally relaxed
minicircle~\cite{skor18}:
\begin{eqnarray}
\Omega_1^{(tr)} &=& \frac{\mu \sin (\omega_0 s)}{A_1} \; ,
\nonumber\\
\Omega_2^{(tr)} &=& \frac{\mu \cos (\omega_0 s)}{\widetilde{A}_2} \; ,
\label{om_tr} \\
\Omega_3^{(tr)} &=& -\frac{\mu G \cos (\omega_0 s)}{C \widetilde{A}_2}  \; .
\nonumber
\end{eqnarray}
In the previous equations ${\mu} \equiv l_b/R$, where $R$ is the average
radius of the circle and where $l_b$, given by~\eqref{lb}, is the bending
persistence length of the model (see Ref.~\onlinecite{skor18} for more details).

Equations~\eqref{om_tr} yield a perfect circle only in the case of $A_1
= \widetilde{A}_2$, corresponding to a constant curvature $\kappa =
\sqrt{\Omega_1^2 + \Omega_2^2}$.  In all other cases the above equations
describe a quasi-circular shape that exhibits small off-planar and
in-planar oscillations, with a modulated total curvature $\kappa$. The
analysis of Eq.~\eqref{om_tr} shows that the dimensionless quantity
$\kappa R$ ($R$ is the average radius of curvature of the minicircle)
is bounded within the interval
\begin{equation}
\frac{\widetilde{A}_2}{A_1+\widetilde{A}_2} 
\leq \frac{\kappa R}{2} \leq 
\frac{A_1}{A_1+\widetilde{A}_2} \; ,
\label{eq:kappa_tr}
\end{equation}
which reduces to a perfect circle $\kappa R=1$ in the isotropic TWLC
limit $A_1=A_2$, $G=0$ (so that $\widetilde{A}_2=A_2$). Note that $A_2$
is the bending stiffness along the easy axis hence $A_2 < A_1$ and
twist-bend coupling enhances the curvature anisotropy since ${\widetilde
A}_2 < A_2$. Twist-bend coupling induces oscillations in the twist,
referred to as "twist waves" in Ref.~\onlinecite{skor18}, which are in
antiphase with the "roll" ($\Omega_2^{(tr)}$) wave. Twist oscillations
are experimentally observed in crystal structures of DNA wrapped around
histone proteins~\cite{rich03}, however their origin has been so far
attributed to an interaction with the underlying histone core proteins,
while Eq.~\eqref{om_tr} shows that these oscillations are a natural
feature of bent DNA, directly deriving from the effect of twist-bend
coupling~\cite{skor18}.

The elastic energy associated with this configuration is obtained by
inserting equations~\eqref{om_tr} into~\eqref{eq:energy}. Using $\mu =
l_b/R$ and $L = 2\pi R$, for a torsionally relaxed minicircle one finds
\begin{equation}
\beta E_\text{tr} = \int_0^L \left[ 
\frac{\mu^2 \sin^2 (\omega_0 s)}{2 A_1} +
\frac{\mu^2 \cos^2 (\omega_0 s)}{2 \widetilde{A}_2} 
\right] \dd s = \frac{\pi l_b}{R} \; ,
\end{equation}
which is formally identical to the energy of a TWLC minicircle.
The difference being that in the TWLC the persistence length $l_b$
is the harmonic mean of the bending stiffnesses, whereas in model
\eqref{eq:energy} the same quantity is a function of all elastic
parameters of the model, see~\eqref{lb} and~\eqref{shiftA2}.

\subsection{Over and undertwisted minicircles}

In the case of torsionally constrained minicircles, one should impose
a constraint of a fixed linking number $Lk$. The White theorem states
that the linking number is the sum of twist and writhe $Lk = Tw + Wr$. As
we are interested in quasi-planar conformations the writhe is small and
$Lk \approx Tw$. Therefore, we impose a constraint on a twist instead
by introducing an additional Lagrange multiplier as follows
\begin{equation}
\beta \widehat{E} \approx
\beta E -  \int_0^L \left[\mu \Omega_1 \sin (\omega s) + 
\mu\Omega_2 \cos (\omega s) + \lambda \Omega_3 \right] \dd s \; ,
\label{lagrange2}
\end{equation}
where we have used the same approximation as in \eqref{lagrange1}
but with ${\widehat{\bf x}} = \sin( \omega s) {\widehat{\bf e}}_1+
\cos (\omega s) {\widehat{\bf e}}_2$, in order to take into account
that the introduction of a constraint in $\Omega_3$ induces a shift in
the intrinsic twist to $\omega \equiv \omega_0 + \Delta \omega$. Local
energy minimization yields the modified equations
\begin{eqnarray}
\Omega_1^{(tc)} &=& \frac{\mu \sin (\omega s)}{A_1} \; ,
\nonumber\\
\Omega_2^{(tc)} &=& \frac{\mu \cos (\omega s)}{\widetilde{A}_2} 
- \frac{\lambda G}{C \widetilde{A}_2} \; ,
\label{om_tc}\\
\Omega_3^{(tc)} &=& -\frac{\mu G \cos (\omega s)}{C\widetilde{A}_2} 
+ \frac{\lambda A_2}{C\widetilde{A}_2} \; .
\nonumber
\end{eqnarray}
As expected, a non-zero $\lambda$ introduces an offset in $\Omega_3$,
i.e.  an average excess twist density given by 
\begin{equation} 
\Delta \omega = \frac{\lambda A_2}{C\widetilde{A}_2}
\end{equation} 
where again we assumed that the writhe has negligible
contribution. Because of twist-bend coupling there is a corresponding
offset in $\Omega_2$ as well, which enhances the inhomogeneity in
the curvature.  The shape described by Eqs.~\eqref{om_tc} alternates
between high and low curvature regions, but the excess twist induces
oscillations in $\kappa R$ in a wider range when compared to the
torsionally relaxed case~\eqref{eq:kappa_tr} (more details are shown in
Appendix~\ref{app:curv}). The $\Omega$'s from Eqs.~\eqref{om_tc} are in
very good agreement with numerical results on coarse-grained DNA models
discussed in the next section.

Similarly to what was done for the untwisted case~\cite{skor18}, one
can show that the Lagrange multiplier $\mu$ fixes the average radius
of curvature, hence $\mu = l_b/R$. Plugging Eqs.~\eqref{om_tc} 
into~\eqref{eq:energy} we obtain the following energy for a torsionally
contrained minicircle
\begin{eqnarray}
\beta E_\text{tc} &=& \int_0^L \left[ 
\frac{\mu^2 \sin^2 (\omega s)}{2 A_1} +
\frac{\mu^2 \cos^2 (\omega s)}{2 \widetilde{A}_2} +
\frac{\widetilde{C}}{2} \Delta \omega^2
\right. \nonumber \\
&-& \left.
\left( \frac{2G}{\widetilde{A}_2} + \frac{G}{A_2} \right) 
\mu \cos (\omega s) \Delta \omega \right] \dd s =
\frac{\pi l_b}{R} + L \frac{\widetilde{C}}{2} \Delta \omega^2
\nonumber\\
\label{eq:en_tc}
\end{eqnarray}
where $\widetilde{C}$ has been defined by Eq.~\eqref{lt} and the term
proportional to $\cos(\omega s)$ averages out in the integration. As in
the untwisted case the energy is formally identical to that of a TWLC.
Again in model~\eqref{eq:energy}, the difference with TWLC is that
the bending stiffness $l_b$ and torsional stiffness $\widetilde{C}$
are function of all elastic constant.  Naturally, for large excess
twist the approximation of negligible writhe will break down and the
minicircle will start supercoiling. This instability is not captured
by the present model~\eqref{lagrange2}, as it was derived by a local
constraint, \revision{explicitly} neglecting this effect.

%%%%%%%%%%%%%%%%%%%%%%%%%%%%%%%%%%%%%%%%%%%%%%%%%%%%%%%%%%%%%%%
%%%%%%%%%%%%%%%%%%%%%%%%%%%%%%%%%%%%%%%%%%%%%%%%%%%%%%%%%%%%%%%
%%%%%%%%%%%%%%%%%%%%%%%%%%%%%%%%%%%%%%%%%%%%%%%%%%%%%%%%%%%%%%%
\section{Numerical Results}
\label{SEC:NUMERICS}

\revision{In order to check the theoretical predictions,
we performed some numerical calculations using two different
models. The first one, which we refer to as the {\em triad model}
(see also Refs.~\onlinecite{skor18,nomi18}) is obtained by a direct
discretization of the continuum model \eqref{eq:energy}. The second
model is oxDNA \cite{ould10}, a coarse-grained model describing DNA as
two intertwined strings of rigid nucleotides.  }
In the triad model the stiffness constants $A_1$, $A_2$, $C$ and $G$
are input parameters and can be freely chosen, while for oxDNA their
values are fixed by the force field parametrization, which was tuned to
reproduce known structural, thermodynamical and mechanical properties
of DNA \cite{ould10}. In addition, oxDNA comes in two versions:
an implementation with symmetric grooves (oxDNA1) and \revision{an
improved} version that \revision{explicitly introduces} asymmetric
grooves (oxDNA2).  This feature makes the model suitable to test the
implications of twist-bend coupling, which derives precisely from the
groove asymmetry \cite{mark94}.  The elastic parameters for oxDNA1
and oxDNA2 were computed in Ref.\onlinecite{skor17} from the analysis
of equilibrium fluctuations of a linear molecule and the results are
shown in Table~\ref{tab1}, yielding in particular a value of $G$ which
is comparable to that of the other elastic constants.

%%%%%%%%%%%%%%%%%%%%%%%%%%%%%%%%%%%%%%%%%%%%%%%%%%%%%%%%%%%%%%%
\setlength{\tabcolsep}{7pt} % Space between columns
\begin{table}[t]
\caption{Values of the stiffness coefficients (in nm) for oxDNA1 and
oxDNA2 obtained in Ref.\onlinecite{skor17}. Note that while oxDNA1 has
negligible twist-bend coupling $G < 0.3$~nm, this coupling is $G=30$~nm
in oxDNA2. The values in the table define two set of parameters of the
triad model referred to as set ${\bf M}^1 = (A_1,A_2,C,G)=(84,29,118,0)$
and set ${\bf M}^2 = (A_1,A_2,C,G)=(81,39,105,30)$. The last
column gives $\widetilde{A}_2$ and $\widetilde{C}$ as obtained from
Eqs.~\eqref{shiftA2} and~\eqref{lt}, respectively.}
\begin{ruledtabular}
\centering
\begin{tabular}{c c c c c | c c}

% \toprule

& $A_1$ & $A_2$ & $C$ & $G$ & $\widetilde{A}_2$ & $\widetilde{C}$\\

% \midrule\\[-10pt]
 \hline %[-6 pt]

oxDNA1  & 84(14) & 29(2) & 118(1) & 0.1(0.2) & 29 & 118\\
oxDNA2  & 81(10) & 39(2) & 105(1) & 30(1) & 30 & 82 %\\[4pt]

\end{tabular}
\label{tab1}
\end{ruledtabular}
\end{table}
%%%%%%%%%%%%%%%%%%%%%%%%%%%%%%%%%%%%%%%%%%%%%%%%%%%%%%%%%%%%%%%

\subsection{MC simulations with the Triad Model}

In the triad model, a double stranded DNA of $N$ base pairs is
represented by $N$ beads, each carrying a frame of three orthogonal unit
vectors $\lbrace \widehat{\bf e}_1, \widehat{\bf e}_2, \widehat{\bf
e}_3 \rbrace$. The distance between consecutive beads is fixed and
equal to $a=0.33$ nm and the vector $\widehat{\bf e}_3$ always points
towards the sequentially adjacent bead. Given two consecutive triads,
the deformation parameters are defined by a definition analogous to
Eq.~\eqref{eq:rotation} valid for finite rigid body rotations (for details
see Supplemental Material of Ref.~\onlinecite{skor17}), while the energy
of a conformation is obtained by discretizing the continuum energy model,
Eq.~\eqref{eq:energy}. Low temperature \revision{Monte Carlo} (MC) simulations 
have been carried out using the two sets of parameters ${\bf M}^1$
and ${\bf M}^2$ matching the oxDNA1 and oxDNA2 values (see caption of
Table~\ref{tab1}).

%%%%%%%%%%%%%%%%%%%%%%%%%%%%%%%%%%%%%%%%%%%%%%%%%%%%%%%%%%%%%%%%%%%%%%%%
\begin{figure*}[t]
\includegraphics[scale=1]{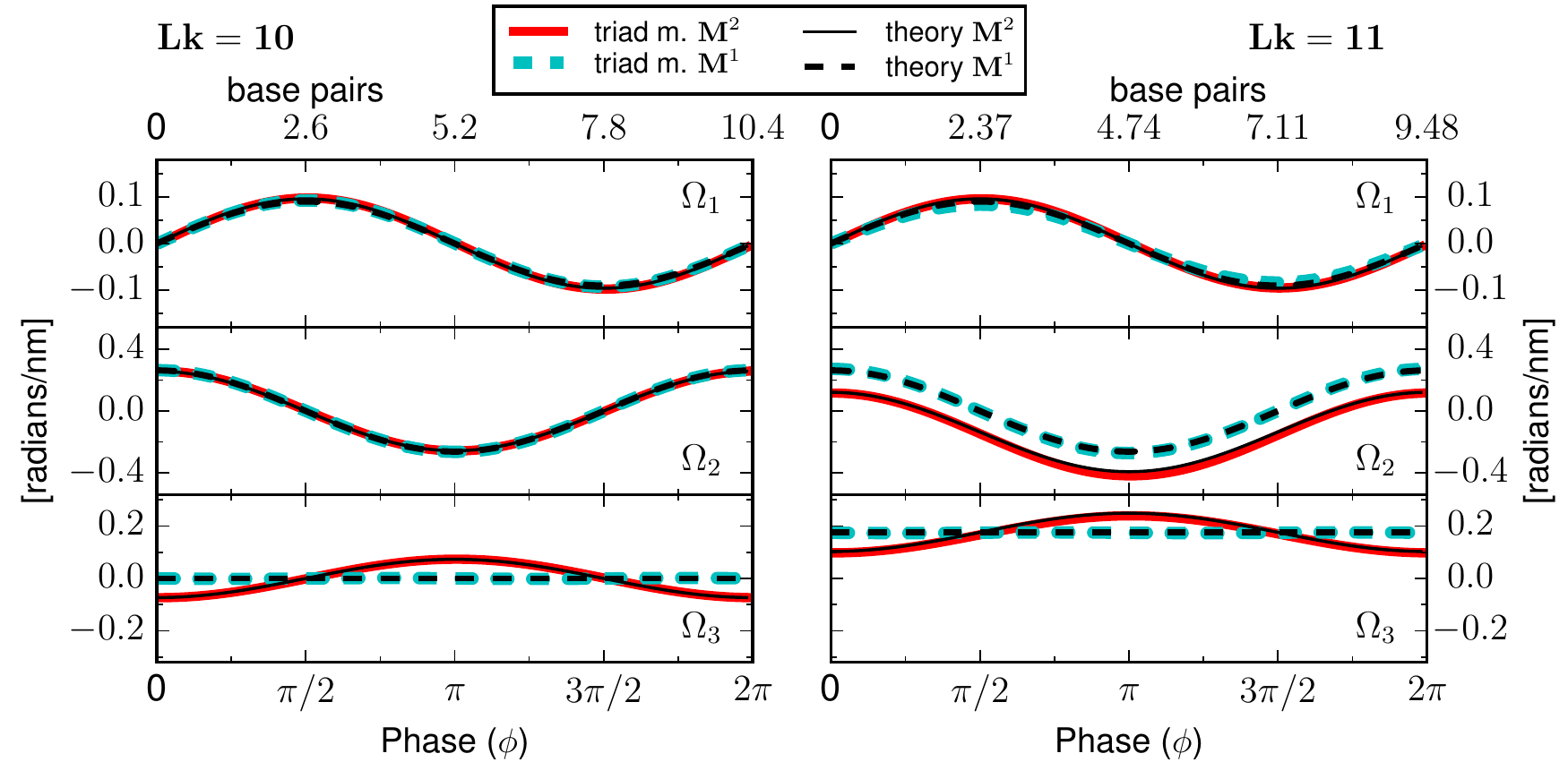}
\caption{\label{FIG:minicirclesTM} Comparison between deformation
parameters ($\Omega_i$'s) of a minicircle of $N=104$ bp obtained from
MC simulations and from analytical results. Left: Torsionally relaxed
minicircle with linking number $Lk = 10$ ($\omega = \omega_0 = 1.85$
nm$^{-1}$). Right panels: Overtwisted minicircle with linking number $Lk
= 11$ ($\omega = 2.03$ nm$^{-1}$). Each single panel reports $\Omega_i$
from MC simulations with stiffness parameters from the set ${\bf M}^1$
(dashed cyan line) and from the set ${\bf M}^2$ (red solid line), as givem
in Table~\ref{tab1} together with the analytical expressions for the same
values of the parameters. MC data for the torsionally relaxed case $Lk=10$
are in excellent agreement with Eqs.~\eqref{om_tr}, while the torsionally
constrained case $Lk=11$ simulations agree with Eqs.~\eqref{om_tc}.
Results of MC simulations are obtained by taking the Fast Fourier
Transformation (FFT) of different configurations, shifting all phases in
such a way that for each MC configuration $\Omega_1 (0) = 0$, taking the
inverse FFT and then averaging the reconstructed signal over the different
configurations. Averages are done on $250$ independent configurations.
The bottom horizontal axis shows the local phase $\omega s$ while the top
horizontal axis shows the polymer coordinate $s$ in terms of base pairs.}
\end{figure*}
%%%%%%%%%%%%%%%%%%%%%%%%%%%%%%%%%%%%%%%%%%%%%%%%%%%%%%%%%%%%%%%%%%%%%%%%

Figure~\ref{FIG:minicirclesTM} shows a comparison of the $\Omega_i$
obtained from the triad model simulations (colored lines) with the
analytical predictions (black lines). In the simulations we fix the
linking number $Lk$, which is a topological invariant measuring the
number of times the two strands are wound around each other. The White
theorem states that the linking number is the sum of twist and writhe
$Lk = Tw + Wr$. As we are interested in quasi-planar conformations
the writhe is small and $Lk \approx Tw$. Therefore the constraint
used in~\eqref{lagrange2} to fix the excess twist density $\Omega_3$
is expected to adequately describe closed circular DNA. The left panel
of Fig.~\ref{FIG:minicirclesTM} shows a torsionally relaxed minicircle
with $104$ base pairs and linking number $Lk=10$, while in the right
panel the circle is overtwisted and has $Lk=11$.  All analyzed cases
display excellent agreement between analytical models and MC data.
The horizontal axis shows a single helical turn, corresponding to a
phase $0 \leq \phi \leq 2\pi$ obtained from the analysis of the Fourier
spectrum of $\Omega_i$, as explained in the caption. In a torsionally
relaxed DNA one helical turn corresponds to $10.4$ base pairs as shown
in the horizontal top scale of the left panel. In the overtwisted case
($Lk=11$) one helical turn corresponds to $9.5$ base pairs.

As shown in Fig.~\ref{FIG:minicirclesTM}, in the torsionally
relaxed case $Lk=10$ and for the set ${\bf M}^2$ ($G \neq 0$) the
twist oscillates, while these oscillations are absent for the set
${\bf M}^1$ ($G=0$). In the torsionally constrained case $Lk=11$
the twist of both sets is shifted, see bottom right graph of
Fig.~\ref{FIG:minicirclesTM}. Overtwisting affects the $\Omega_2$
of set ${\bf M}^2$, but not that of set ${\bf M}^1$, as predicted
by the analytical model. Is is worth emphasizing that the theoretical
predictions of both panels do not contain adjustable parameters.  In fact,
the Lagrange multipliers $\mu$ and $\lambda$ are fixed as follows: $\mu
= l_b/R$, where $R$ is the radius of a perfect circular chain of $104$
beads with a distance between consecutive beads of $0.33$ nm ($R \simeq
5.46$ nm), $\lambda = C \tilde{A}_2 \Delta \omega / A_2$ with $\Delta
\omega = 0.18$ nm$^{-1}$, and the stiffness parameters are given. In the
undertwisted case (not shown) very similar profiles for the $\Omega_i$
are observed, but the shifts in $\Omega_2$ and $\Omega_3$ carry the
opposite sign.

%%%%%%%%%%%%%%%%%%%%%%%%%%%%%%%%%%%%%%%%%%%%%%%%%%%%%%%%%%%%%%%%%%%%%%%%
\begin{figure*}[t]
\includegraphics[scale=1]{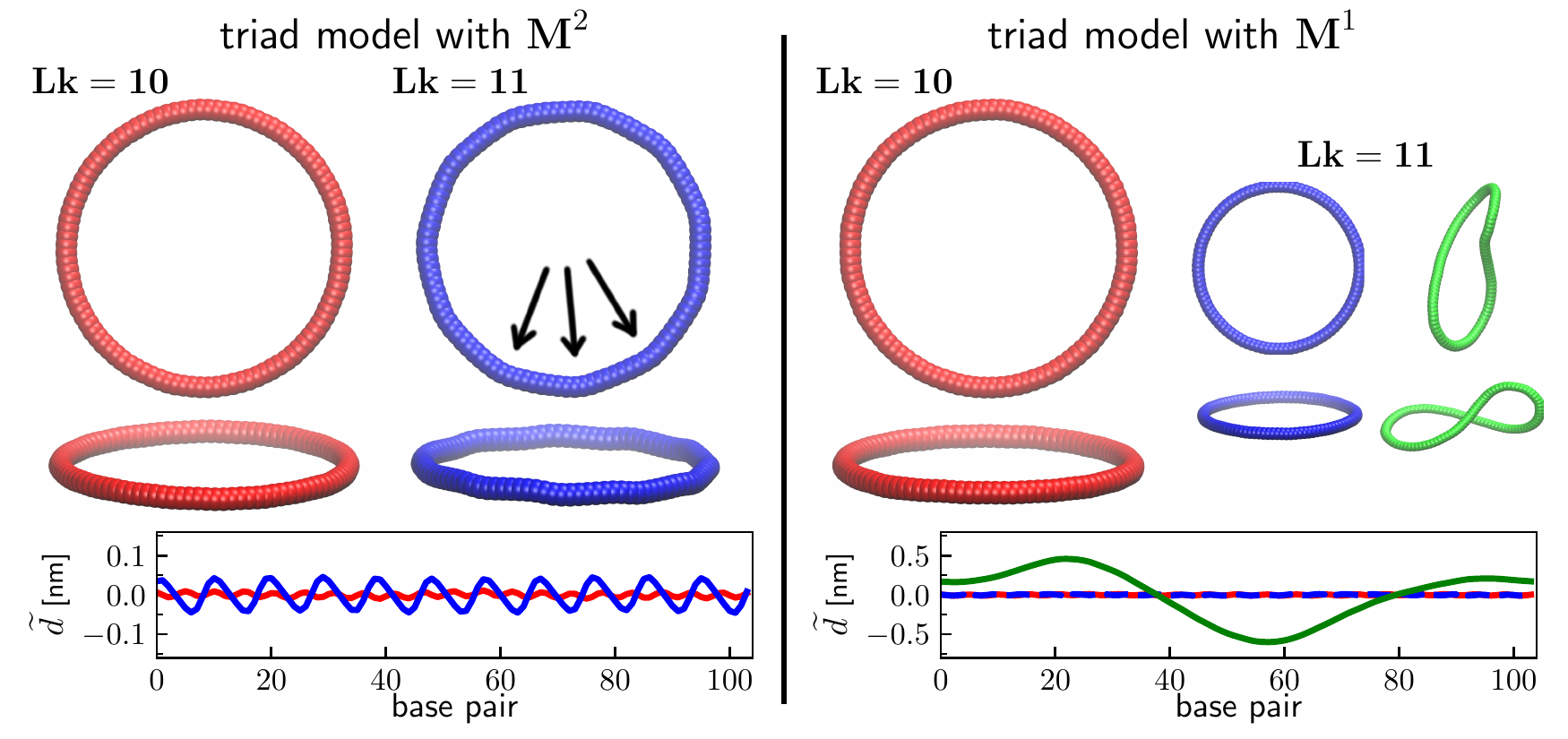} 
\caption{\label{FIG:shapesTM} Typical shapes obtained from low temperature
MC simulations of a minicircle of $N=104$ bp modeled with the triad model.
Minicircle are relaxed ($Lk=10$) or overtwisted ($Lk=11$). Left panel:
stiffness parameters from set ${\bf M}^2$, which has a non-zero twist-bend
coupling. \revision{The black arrows point at three vertices of the
polygon.} Right panel: stiffness parameters ${\bf M}^1$. The bottom graph
of each panel reports the signed distance $\widetilde{d}=(ax_1 + by_1 +
z_1 + c)/\sqrt{a^2 + b^2 + 1}$ (where $ax + by + z + c=0$ is the plane
equation and $(x_1,y_1,z_1)$ the bead coordinates) of the beads from the
plane that best fit the beads positions. Colors are consistent with the
snapshots shown in the panel.}
\end{figure*}
%%%%%%%%%%%%%%%%%%%%%%%%%%%%%%%%%%%%%%%%%%%%%%%%%%%%%%%%%%%%%%%%%%%%%%%%

Figure~\ref{FIG:shapesTM} shows the typical shapes of relaxed and
overtwisted minicircles. In both sets ${\bf M}^1$ and ${\bf M}^2$, the
relaxed configuration ($Lk=10$) is almost completely planar and closely
resembles a perfect circle.  However, the introduction of additional
twist leads to remarkable differences in the behavior of minicircles
parameterized by the sets ${\bf M}^1$ and ${\bf M}^2$. The latter starts
to exhibit the shape of a rounded polygon or, more precisely, a rounded
hendecagon, where the amount of vertices is induced by the imposed excess
linking number.  Furthermore, the structure is moderately off-planar,
as illustrated by the plot of $\tilde{d}$, the signed distance of each
base pair from the best fitting plane vs. base pair position~\cite{note2}
(see left lower panel of Fig.~\ref{FIG:shapesTM}).  Consistent with the
oscillations of the strain fields $\Omega_i$, $\tilde{d}$ fluctuates
with a wavelength of $2\pi/\omega$.  On the other hand, when one
considers the parameter set without twist-bend coupling (${\bf M}^1$),
depending on the setup of the MC simulation, two typical configurations
are found shown in blue and green in Fig.~\ref{FIG:shapesTM}, right. If
one starts from a perfectly circular overtwisted shape and performs MC
updates at low temperature, the simulation relaxes to the shape drawn in
blue (Fig.~\ref{FIG:shapesTM}, right). From this shape the $\Omega_i$
shown in Fig.~\ref{FIG:minicirclesTM} were calculated. If, on the
other side, one starts from a high temperature simulation and gradually
lowers the temperature to reach the ground state a strongly off planar
conformation, as that shown in green in Fig.~\ref{FIG:minicirclesTM}, is
obtained. The latter shows an onset of supercoiling, which is not found
in the simulations with set ${\bf M}^2$ \revision{for the same value
of $Lk$}. In that case the polygon shape is always recovered at low
temperatures, irrespectively from the simulation path followed. This shows
that the model ${\bf M}^1$ is more prone to supercoiling compared to model
${\bf M}^2$, for which the torsional stress is released in bending and
off-planar fluctuations. This is also reflected in Eq.~\eqref{eq:en_tc},
which shows that the torsional energy is controlled by the parameter
$\widetilde{C}$~\eqref{lt}, rather than the intrinsic twist stiffness
$C$. The sets ${\bf M}^1$ and ${\bf M}^2$ have comparable twist
stiffness $C$, but $\widetilde{C}=C=118$~nm for set ${\bf M}^1$ as
$G=0$, while $\widetilde{C}=82$~nm and $C=105$~nm for set ${\bf M}^2$
which implies a considerable lower twisting energy for the same amount
of overtwisting. This explains why model ${\bf M}^1$ supercoils more
easily when compared to ${\bf M}^2$. \revision{Also the model ${\bf M}^2$
eventually supercoils at larger $|\Delta Lk|$ (not shown).  We will not
discuss the properties of the supercoiling transition for the two models
here, which would require sampling both systems at experimentally relevant
temperatures.  Appendix~\ref{app:WrTrajectories} provides some details
of MC simulations with triad model, confirming that model ${\bf M}^2$
has a lower propensity to supercoiling compared to ${\bf M}^1$ once the
same external parameters as linking number and temperatures are chosen.}

\subsection{MD simulations with oxDNA}

Double helical coarse-grained models have become
very popular in the recent few years to study a large
number of equilibrium and dynamical properties of DNA
\cite{ould10,sulc12,fosa16,fred14,lequ17,li18,chak18,coro18}.
oxDNA~\cite{ould10} provides an effective mesoscopic description of
DNA in which each nucleotide is represented by a \revision{rigid} object equipped
with three interactions sites for base pairing, coaxial stacking,
electrostatic and steric interactions, that was tuned to reproduce the
properties of dsDNA.  Langevin dynamics of the system was integrated at
low temperature ($15$ K) with the {\small LAMMPS} package~\cite{plim95}
using the implementation of Henrich et al.~\cite{Henrich2018} and default
values for the interaction parameters.

%%%%%%%%%%%%%%%%%%%%%%%%%%%%%%%%%%%%%%%%%%%%%%%%%%%%%%%%%%%%%%%%%%%%%%%%
\begin{figure*}[t]
\includegraphics[scale=1]{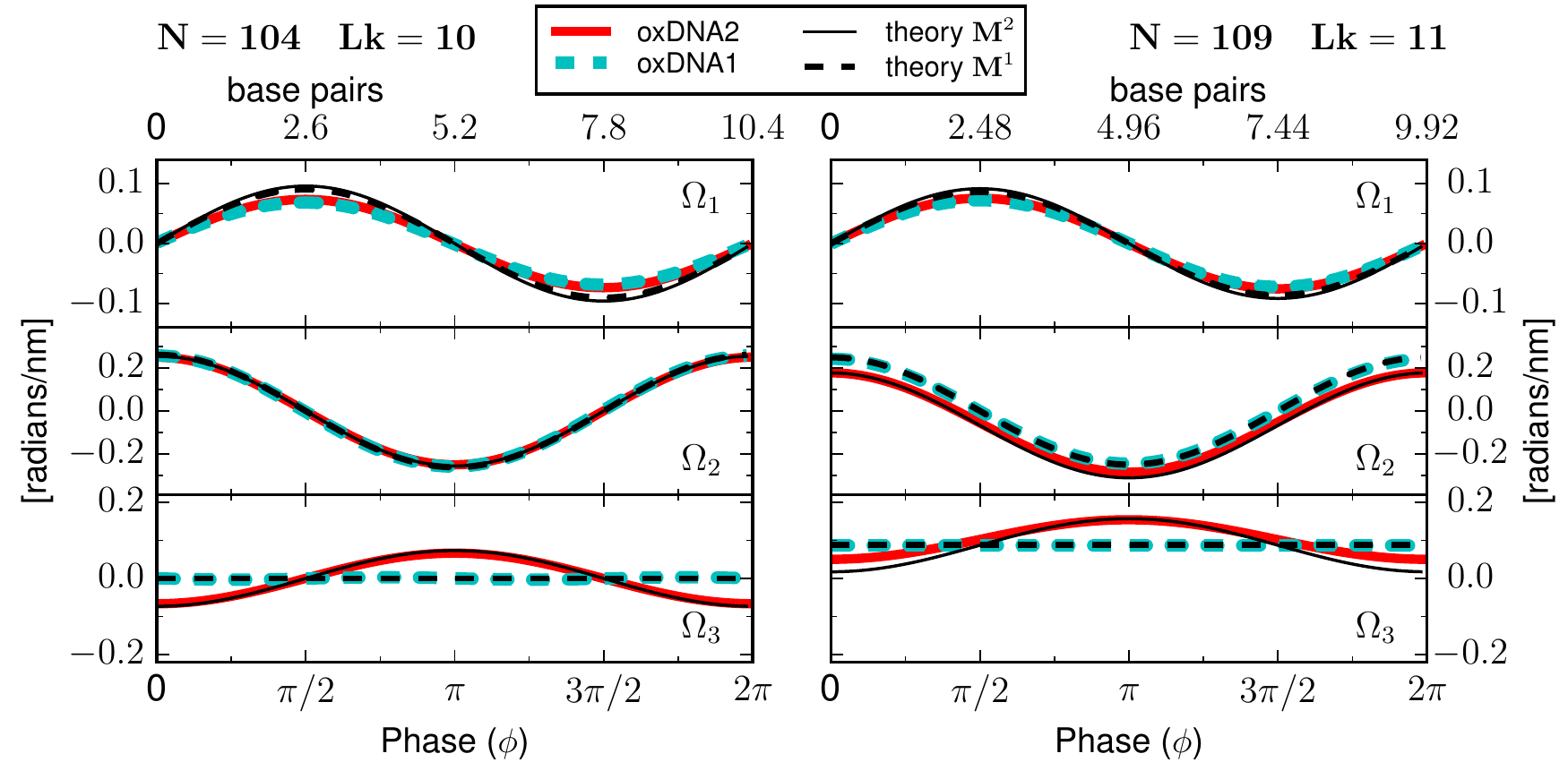} 
\caption{\label{FIG:minicirclesOxDNA} Comparison between the analytical
prediction of deformation parameters ($\Omega_i$'s) and those obtained
from oxDNA MD simulation at $15$ K. Left: Relaxed minicircle of $N=104$
bp and linking number ($Lk$) equal to $10$ ($\omega = \omega_0 =
1.85$ nm$^{-1}$). Right panels: Overtwisted minicircle of $N=109$ bp
with $Lk=11$ ($\omega = 1.93$ nm$^{-1}$). Each single panel reports:
$\Omega_i$ from oxDNA1 (dashed cyan line) and oxDNA2 (red solid line);
analytical expression with stiffness parameters ${\bf M}^1$ (dashed
black line, Eq.~\eqref{om_tr}) and with stiffness parameters ${\bf M}^2$
(solid black line, Eq.~\eqref{om_tc}).  Results of MD simulations
are obtained by using the same procedure described in the caption of
Fig.~\ref{FIG:minicirclesTM}.}
\end{figure*}
%%%%%%%%%%%%%%%%%%%%%%%%%%%%%%%%%%%%%%%%%%%%%%%%%%%%%%%%%%%%%%%%%%%%%%%%

Again, the $\Omega_i$ calculated from oxDNA MD simulations are
consistent with those predicted by the analytical model (see
Fig.~\ref{FIG:minicirclesOxDNA}).  As already pointed out in
Ref.~\onlinecite{skor18}, small deviations can be noticed in the
amplitude of $\Omega_1$ oscillations. Interesting enough, also the
value of $\Omega_3$ for oxDNA2 slightly deviates from the analytical
curve but only in the case in which the minicircle is overtwisted. As
for $\Omega_1$, these small deviations probably arise from some
additional interactions (e.g. higher-order terms) present in oxDNA,
but not considered in the energy functional.  Notice that we considered
an overtwisted minicircle with small additional twist $\Delta \omega /
\omega_0 \simeq 5$\%. Such choice is due to the fact that for higher
additional twist, e.g. of the order of $10$\% as imposed in the case of
Fig.~\ref{FIG:minicirclesTM}, both oxDNA1 and oxDNA2 show a behavior of
strong deviations from planarity and a propensity to form supercoils.

%%%%%%%%%%%%%%%%%%%%%%%%%%%%%%%%%%%%%%%%%%%%%%%%%%%%%%%%%%%%%%%%%%%%%%%%
\begin{figure*}[t]
\includegraphics[scale=1]{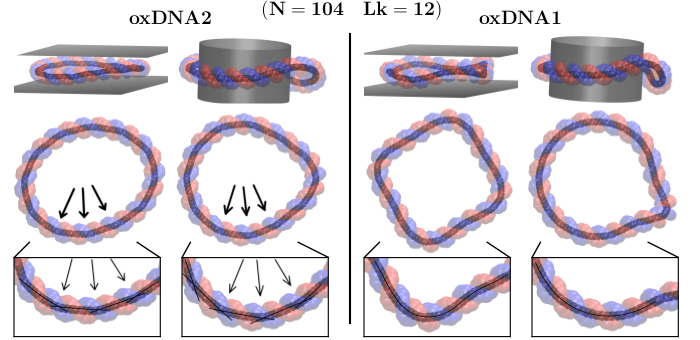} 
\caption{\label{FIG:shapesOxDNA} Typical shapes of an overtwisted
minicircles of $104$ bp and $Lk = 12$ ($20\%$ excess linking number
fraction $\Delta Lk/Lk_0 = 0.2$). Shapes are from MD simulations at
$15$ K of oxDNA1 (right panel) and oxDNA2 (left panel) and minicircles
are confined into a slit or outside a cylindrical surface. Blue and red
transparent chains represent the two strands while the black solid beads
are the center of mass of each base pair. The distance between the two
planes of the slit is $3.4$ nm while the cylinder radius is $4.4$ nm.
\revision{In the case of oxDNA2, the black arrows point at three vertices
of the polygon. These vertices follow the periodicity of the double helix.
In the case of oxDNA1 confined into a slit, the apparent strong curvature
in the four corners is an artifact of the 2D projection.}}
\end{figure*}
%%%%%%%%%%%%%%%%%%%%%%%%%%%%%%%%%%%%%%%%%%%%%%%%%%%%%%%%%%%%%%%%%%%%%%%%

Next, we considered strongly overtwisted minicircles in confined
geometries so that the DNA cannot easily supercoil. Two cases
were analyzed: a DNA minicircle confined between two flat
surfaces and a minicircle wrapped around a cylinder, shown in
Figure~\ref{FIG:shapesOxDNA}. These two situations can be relevant,
respectively, in the case of AFM experiments where DNA is confined in
2D by absorbption on a mica surface \cite{rive96,vall05,wigg06,vand13}
or in the case of nucleosomes where DNA is wrapped around the
cylindrically shaped octamer of histone proteins \cite{esla16,rich03}.
Figure~\ref{FIG:shapesOxDNA} shows the typical shapes of confined
oxDNA1 and oxDNA2 minicircles when an excess linking number of $20$\%
is introduced. Apart obviously from the twist-bend coupling $G$, oxDNA1
and oxDNA2 have similar elastic parameters (see Table~\ref{tab1}) but
their response to overtwisting and confinement is very different. This
is in agreement to what was found in the triad model: as for the set
${\bf M}_1$, also oxDNA1 has a stronger tendency to develop off-planar
conformation, indicating that it is easier to supercoil. Furthermore, in
the case of oxDNA2, a closer inspection of the base pairs' center of mass,
as shown in the inset of Fig.~\ref{FIG:shapesOxDNA} (left), indicates that
the minicircle tends to have a rounded polygonal shape, \revision{with the
periodicity of the double helix, in agreement with the theory discussed
in this paper}. On the other hand, oxDNA1 has a smoother curvature even
if, under planar confinement, the shape of oxDNA1, once projected in 2D,
resembles a rounded square. 
\revision{The origin of these seemingly strong curved regions is the onset of
a buckling transition, i.e. the apparent strong curvature in the four
corners of the 2D projection is due to writhe rather than curvature
and does not exhibit the periodicity of the inert double helical repeat
length.}

%%%%%%%%%%%%%%%%%%%%%%%%%%%%%%%%%%%%%%%%%%%%%%%%%%%%%%%%%%%%%%%
%%%%%%%%%%%%%%%%%%%%%%%%%%%%%%%%%%%%%%%%%%%%%%%%%%%%%%%%%%%%%%%
%%%%%%%%%%%%%%%%%%%%%%%%%%%%%%%%%%%%%%%%%%%%%%%%%%%%%%%%%%%%%%%
\section{Conclusion}

In this paper we have studied minimal energy shapes of torsionally
constrained circular DNA molecules. As shown earlier~\cite{skor18},
the effect of twist-bend coupling is to produce shapes characterized by
coupled oscillations in twist ($\Omega_3$) and in the bend ($\Omega_2$)
densities. 
\revision{We have extended here the investigation to the effect of a
torsional strain which} induces a net shift in the twist density
$\Omega_3$, forcing it to oscillate around a non-zero average value. As
a consequence of twist-bend coupling this effect is transmitted to
the groove-bending strain $\Omega_2$. The breaking of the symmetry of
$\Omega_2$ oscillations results in a shape resembling that of a rounded
regular polygon with a periodic alternation of high and low curvature
regions. We have shown (extending the theory of Ref.~\onlinecite{skor18})
that a simple analytical model reproduces very well the shapes obtained
from simulations. The comparison between theory and simulations is
remarkable as there are no adjustable parameters.

The analytical model provides a simple way to estimate the energy
of torsionally relaxed and torsionally constrained minicircles. It
turns out that, due to the peculiar shapes of the circles induced by
twist-bend coupling, 
\revision{the elastic energy due to bending and twist is} 
not governed by the intrinsic stiffnesses, but by rescaled parameters
$l_b$ and $\widetilde{C}$, given by Eq.~\eqref{lb} and~\eqref{lt}
respectively. As a consequence, although the two sets of parameters
used in this work have comparable torsional stiffness $C$ (see
Table.~\ref{tab1}), their torsional response is very different. In the
set ${\bf M}^1$, with vanishing twist bend coupling, the twist energy
is $\frac{1}{2} C L \Delta \omega^2$ while in ${\bf M}^2$ this is
reduced to $\frac{1}{2} \widetilde{C} L \Delta \omega^2$ (recall that
$\widetilde{C}<C$, see~\eqref{lt}). In the set ${\bf M}^2$ 
the minimal energy shape
\revision{exploits} the presence of a cross term $G \Omega_2 \Omega_3$
which can become negative, \revision{hence lowering the energy}, if
$\Omega_2$ and $\Omega_3$ have opposite signs. The %appropriate 
value
of torsional elastic constant for DNA has been %debated, 
\revision{discussed at length in the literature} with different
techniques~\cite{horo84,fuji90,volo97,moro98,bouc98,kaue11,brya12}
providing values ranging typically from $C=75$~nm to $C=110$~nm,
\revision{although occasionally lower or higher values have been
reported (a table collecting the elastic constant measurements reported
in the literature from various experimental techniques can be found in
the supplemental of Ref.~\onlinecite{nomi17})}. One way of extracting
$C$ is from the analysis of dynamical or equilibrium properties of DNA
minicircles, see eg. Refs.~\onlinecite{shor83a,horo84}. \revision{In this
analysis it is assumed that DNA is described by an elastic rod model with
independent twist and bending deformations (TWLC) and that if the circles
are sufficiently small ($\lesssim 200$~bp) the contribution from writhe
fluctuations can be neglected \cite{horo84}. Measurements of topoisomers,
i.e. sequences of equal length that differ only by their linking number,
have been used to estimate the bare torsional stiffness $C$. We have shown
here that, in presence of twist-bend coupling, the energetic behavior
of minicircles is governed by a renormalized stiffness $\widetilde{C}$,
which is smaller than $C$ and contains the parameters $G$ and $A_2$,
see~\eqref{lt}.}

\revision{In a recent paper~\cite{skor18} some of us showed that minimal
energy shapes of minicircles with twist-bend coupling fit well structural
nucleosomal DNA data, as obtained from X-ray crystallography. Nucleosomal
DNA is wrapped around the nucleosome, a stable complex formed by histone
proteins tightly bound to each other. The nucleosome is known to slide
along the DNA and one of the most discussed mechanisms is that of the
diffusion of twist defects, see Ref.~\onlinecite{esla16} for a recent
review. We have shown here that local under or overtwisting of bent
DNA is accompanied by a change in shape with a modulation of the local
curvature, which may influence the way the twist defects propagate
along the nucleosomal DNA sequence. This is an interesting issue to be
considered in future work.}

%%%%%%%%%%%%%%%%%%%%%%%%%%%%%%%%%%%%%%%%%%%%%%%%%%%%%%%%%%%%%%%
%%%%%%%%%%%%%%%%%%%%%%%%%%%%%%%%%%%%%%%%%%%%%%%%%%%%%%%%%%%%%%%
%%%%%%%%%%%%%%%%%%%%%%%%%%%%%%%%%%%%%%%%%%%%%%%%%%%%%%%%%%%%%%%

\acknowledgments{Discussions with M.\ Laleman, J.\ Marko, S.\ Nomidis and
J.M.\ Schurr are gratefully acknowledged. MC and ES aknowledge financial
support from KU Leuven grant C12/17/006.}

%%%%%%%%%%%%%%%%%%%%%%%%%%%%%%%%%%%%%%%%%%%%%%%%%%%%%%%%%%%%%%%
%%%%%%%%%%%%%%%%%%%%%%%%%%%%%%%%%%%%%%%%%%%%%%%%%%%%%%%%%%%%%%%
%%%%%%%%%%%%%%%%%%%%%%%%%%%%%%%%%%%%%%%%%%%%%%%%%%%%%%%%%%%%%%%

\appendix

\section{Curvature}
\label{app:curv}

From Eqs.~\eqref{om_tc} and using $\mu = l_b/R$ one finds
\begin{eqnarray}
\frac{\kappa^2 R^2}{4} = \frac{\widetilde{A}_2^2}{(A_1+\widetilde{A}_2)^2}
+ \Gamma^2 + \frac{A_1 - \widetilde{A}_2}{A_1+\widetilde{A}_2} \, \cos^2
(\omega s) - \frac{2 \Gamma {A}_1 \cos (\omega s)}{A_1+\widetilde{A}_2}
\; , \nonumber \\ \,
\label{app:kR}
\end{eqnarray}
where we defined
\begin{equation}
\Gamma \equiv \frac{G}{2 A_2} \, R \Delta \omega
\approx \frac{G}{2 A_2} \, \Delta Lk 
\end{equation}
($\Gamma$ is a rescaled dimensionless excess twist density, or
equivalently the \revision{excess} linking number if one neglects 
the contribution of the writhe $2\pi R \Delta \omega \approx  \Delta Lk$).
The torsionally relaxed case  corresponds to $\Gamma=\Delta \omega =
0$, $\omega=\omega_0$. Equation~\eqref{app:kR} gives in this case
a maximal curvature when $\cos (\omega_0 s) = \pm 1$ and a minimal
curvature when $\cos (\omega_0 s) = 0$. These are the bounds given in
Eq.~\eqref{eq:kappa_tr}.

In the torsionally constrained case and for non vanishing twist-bend
coupling one has $\Gamma \neq 0$. The analysis of \eqref{app:kR} yields
the following bounds
\begin{equation}
\frac{\widetilde{A}_2}{A_1+\widetilde{A}_2}
\left(1 - 
\frac{A_1 + \widetilde{A}_2}{A_1-\widetilde{A}_2} 
\, \Gamma^2 \right)^{1/2} 
\leq \frac{\kappa R}{2} \leq
\frac{A_1}{A_1+\widetilde{A}_2} + |\Gamma|
\label{eq:kappa_tc}
\end{equation}
valid for 
\begin{equation}
|\Gamma| \leq \Gamma^* \equiv \frac{A_1 - \widetilde{A}_2}{A_1} 
\end{equation}
and
\begin{equation}
\frac{A_1}{A_1 + \widetilde{A}_2} - |\Gamma|
\leq \frac{\kappa R}{2} \leq
\frac{A_1}{A_1 + \widetilde{A}_2} + |\Gamma|
\label{eq:kappa_tc2}
\end{equation}
for $|\Gamma| \geq \Gamma^*$. In the limit $\Gamma=0$,
Eq.~\eqref{eq:kappa_tc} reduces to \eqref{eq:kappa_tr}.  Comparing
\eqref{eq:kappa_tc} and \eqref{eq:kappa_tc2} with \eqref{eq:kappa_tr}
one sees that introducing an excess twist indeed increases the range of
values through which the curvature $\kappa$ oscillates.

%%%%%%%%%%%%%%%%%%%%%%%%%%%%%%%%%%%%%%%%%%%%%%%%%%%%%%%%%%%%%%%%%%%%%%%%
\begin{figure*}[t]
\includegraphics[scale=1]{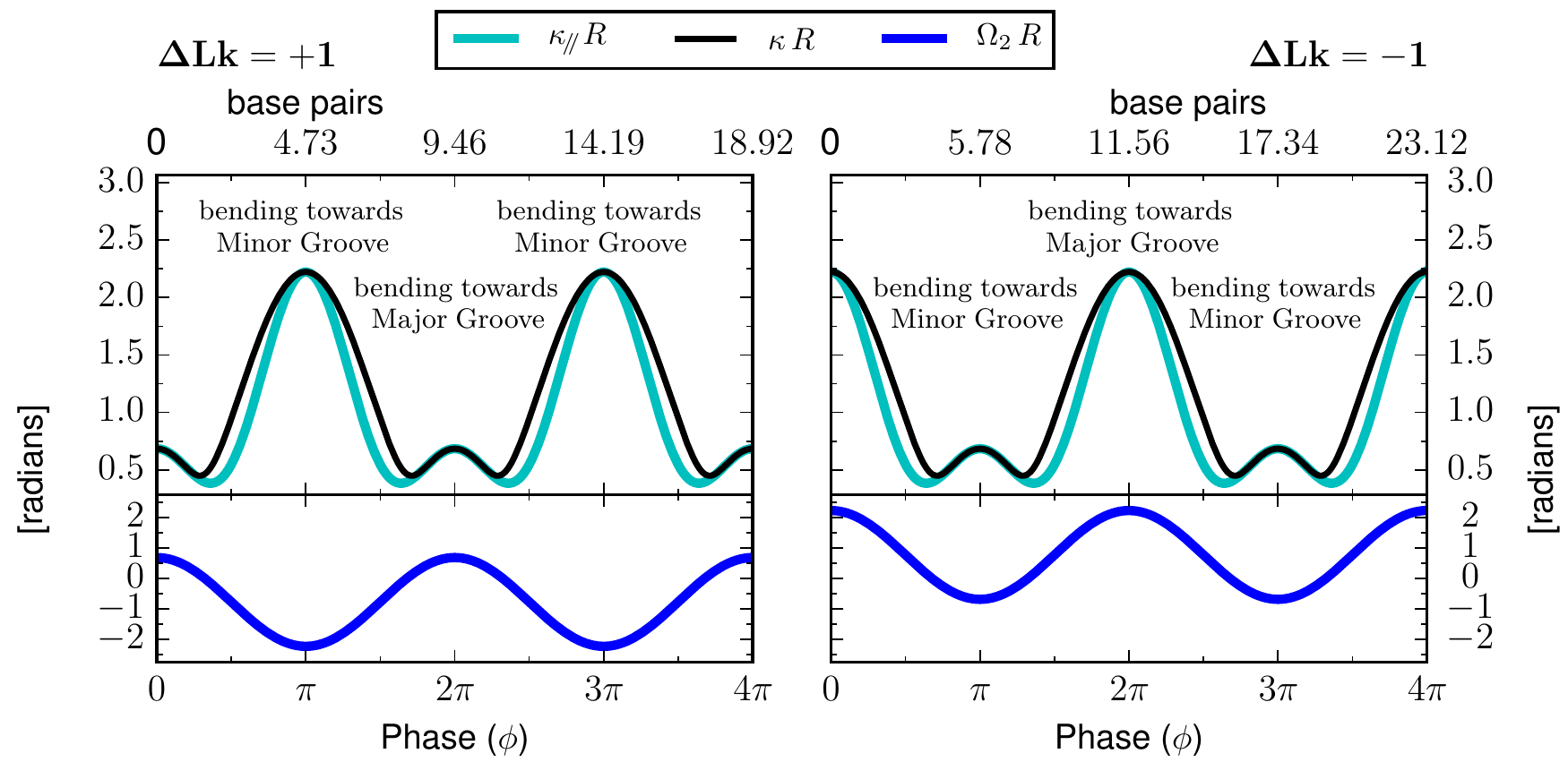} 
\caption{Top panels: Plots of total curvature $\kappa$ and in-plane
curvature $\kappa_{\mathbin{\!/\mkern-4mu/\!}}$ for overtwisted
and undertwisted minicircles.  These plots are obtained from
Eqs.~\eqref{app:kR} and \eqref{app:k_par}, respectively, using the
stiffness parameters ${\bf M}^2 = (A_1,A_2,C,G)=(81,39,105,30)$ The
torsional contraint and twist-bend coupling enhances curvature
anisotropy. Lower panels: Plots of $\Omega_2$ as given by
Eqs.~\eqref{om_tc} A positive value of $\Omega_2$ corresponds to a
bending towards the major groove.}
\label{FIG:curvature}
\end{figure*}
%%%%%%%%%%%%%%%%%%%%%%%%%%%%%%%%%%%%%%%%%%%%%%%%%%%%%%%%%%%%%%%%%%%%%%%%

To gain some more insigths on the different contributions
to $\kappa$ one can decompose it into an in-planar component
$\kappa_{\mathbin{\!/\mkern-4mu/\!}} = \vec{\Omega}_b \cdot
{\widehat{\bf x}}$ and an off-planar component $\kappa_{\perp} =
\vec{\Omega}_b \cdot [ {\widehat{\bf e}}_3 \times {\widehat{\bf
x}}]$, where $\widehat{\bf x}$ identifies the plane where the
minicircle lies (see Fig.~\ref{FIG:intro}). Obviously $\kappa^2 =
\kappa_{\mathbin{\!/\mkern-4mu/\!}}^2 + \kappa_{\perp}^2$. We recall the
definition of bending vector $\vec{\Omega}_b = \Omega_1 \widehat{\bf e}_1
+ \Omega_2 \widehat{\bf e}_2$. Using the approximation of the main text
$\widehat{\bf x} = \sin(\omega s) \, {\widehat{\bf e}}_1 + \cos(\omega s)
\, {\widehat{\bf e}}_2$ one finds:
\begin{eqnarray}
\kappa_{\mathbin{\!/\mkern-4mu/\!}} \, R
&=& 1 + \frac{ A_1 - \widetilde{A}_2 }{A_1 + \widetilde{A}_2 }
\,  \cos\left(2 \omega s\right) - 
2 \Gamma \cos\left(\omega s\right) 
\label{app:k_par}\\
\kappa_{\perp} \,  R
&=& \frac{A_1 - \widetilde{A}_2 }{A_1 + \widetilde{A}_2 } 
\sin (2 \omega s) + 2 \Gamma  \sin (\omega s) 
\label{app:k_perp}
\end{eqnarray}
The top panels of Fig.~\ref{FIG:curvature} show the total curvature
$\kappa$ and in-plane curvature $\kappa_{\mathbin{\!/\mkern-4mu/\!}}$
as given by \eqref{app:kR} and \eqref{app:k_par}, respectively. There
is a small difference between the two, showing that the contribution of
off-planar bending to the total curvature, given by $\kappa_{\perp}$,
is small. For an isotropic model ($A_1 = \widetilde{A}_2$)
with $\Gamma =0$ (corresponding to either $\Delta Lk = 0$
or $G=0$) one recovers from \eqref{app:kR}, \eqref{app:k_par}
and \eqref{app:k_perp} a perfect homogeneous and planar circle
of radius $R$: $\kappa=\kappa_{\mathbin{\!/\mkern-4mu/\!}}=1/R$
and $\kappa_{\perp}=0$. The introduction of anisotropic bending
leads to curvature oscillations, due to the terms proportional
to $\cos(2 \omega s)$ and $\sin(2 \omega s)$ in \eqref{app:k_par}
and \eqref{app:k_perp}. The period of these oscillations is half a
helical repeat.  Overtwisting or undertwisting minicircles with non-zero
twist-bend coupling ($\Gamma \neq0$) breaks this symmetry by introducing
terms that oscillate with a period of a full helical repeat length. The
difference between undertwisting ($\Gamma < 0$) and overtwisting ($\Gamma
> 0$) is that in the former case the region of maximal curvatures $\kappa$
or $\kappa_{\mathbin{\!/\mkern-4mu/\!}}$ corresponds to a global maximum
of $\Omega_2$ ($> 0$) while in the latter to a global minimum, see lower
panels of Fig.~\ref{FIG:curvature}.  The different signs of $\Omega_2$
correspond to different mode of groove bendings\revision{, 
with a positive value of $\Omega_2$ corresponding to
a bending towards the major groove.}

%%%%%%%%%%%%%%%%%%%%%%%%%%%%%%%%%%%%%%%%%%%%%%%%%%%%%%%%%%%%%%%
%%%%%%%%%%%%%%%%%%%%%%%%%%%%%%%%%%%%%%%%%%%%%%%%%%%%%%%%%%%%%%%
%%%%%%%%%%%%%%%%%%%%%%%%%%%%%%%%%%%%%%%%%%%%%%%%%%%%%%%%%%%%%%%

\section{Writhe behavior in MC simulations with triad model}
\label{app:WrTrajectories}

\begin{figure}[t]
\includegraphics[scale=1]{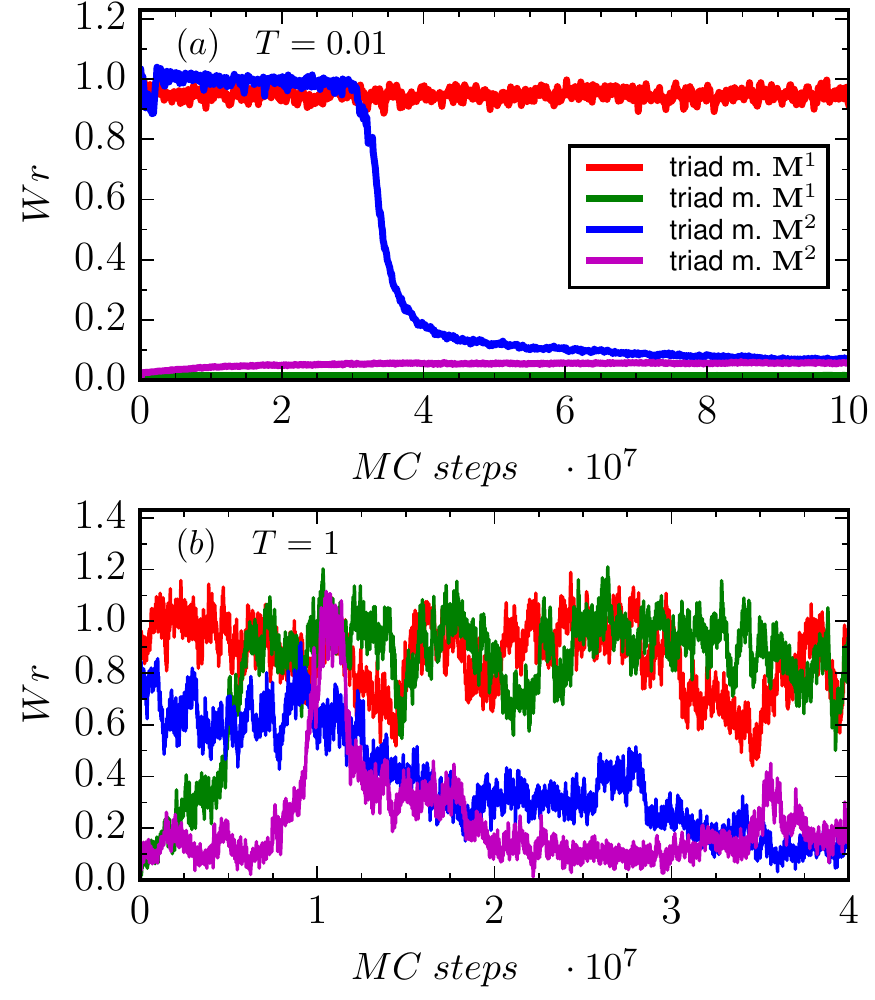} 
\caption{\revision{Plot of writhe, $Wr$, in MC simulations of the
triad model vs. MC staeps at (a) $T=10^{-2}$ and (b) $T=1$ for
overtwisted minicircles with $N=104$ bp and $Lk=11$. Two different
initial conditions are used: perfect overtwisted cirle, with $Wr=0$ 
and supercoiled circle with $Wr=1$.}
}
\label{FIG:WrTrajectories}
\end{figure}

\revision{Figure~\ref{FIG:WrTrajectories} shows plots of the writhe, $Wr$,
as a function of the MC time steps for the triad model. The data are
for overtwisted minicircles with $N=104$ bp and $Lk=11$, obtained using
the same parametrizations considered in Fig.~\ref{FIG:shapesTM} and
complement the results presented that figure. Two different temperature
runs are shown $T=10^{-2}$ (Fig.~\ref{FIG:WrTrajectories}(a)) and $T=1$
(Fig.~\ref{FIG:WrTrajectories}(b)). In this scale $T=1$ corresponds to
room temperature. The simulations are performed for both sets ${\bf M}^1$
and ${\bf M}^2$
with two different initial corresponding to a perfect circular
shape with $Wr=0$ and a supercoiled conformation with $Wr \simeq 1$.
At $T=10^{-2}$ and in the set ${\bf M}^1$ the dynamics does not change
significantly the writhe, which slightly increase for the circular initial
condition to $Wr \simeq 0.02$. This state corresponds to the circular
shape shown in Fig.~\ref{FIG:shapesTM}, right (blue circle $Lk=11$,
set ${\bf M}^1$). The writhe is roughly constant also when starting
from the supercoiled initial condition corresponding to the strongly
off-planar shape of Fig.~\ref{FIG:shapesTM}, right (green $Lk=11$, set
${\bf M}^1$). The low $T$ simulations for ${\bf M}^1$ hence show that
there are two local free energy minima corresponding to the planar
and supercoiled state. For the set ${\bf M}^2$
and $T=10^{-2}$ both initial conditions converge to $Wr \simeq
0.05$, corresponding to the polygonal shape discussed earlier
Fig.~\ref{FIG:shapesTM}, left. Note that the simulation remains for long
time in the supercoiled state before relaxing to the polygonal shape,
indicating that a supercoiled state is metastable.
The writhe is higher than that of the
circle of ${\bf M}^1$, in accordance with the out of plane oscillations
presented in Fig.~\ref{FIG:shapesTM}. 
The analysis was extended to runs at room temperature $T=1$
(Fig.~\ref{FIG:WrTrajectories}(b)). In this case fluctuations in the
shapes are higher. For the set ${\bf M}^1$ both initial conditions
converge to the supercoiled state, whereas for ${\bf M}^2$ a low
writhe conformation is reached. Overall, the results confirm the
low propensity towards supercoiling in the case with a non-vanishing
twist-bend coupling.}

%%%%%%%%%%%%%%%%%%%%%%%%%%%%%%%%%%%%%%%%%%%%%%%%%%%%%%%%%%%%%%%%%%%%%%%%
%%%%%%%%%%%%%%%%%%%%%%%%%%%%%%%%%%%%%%%%%%%%%%%%%%%%%%%%%%%%%%%%%%%%%%%%
%%%%%%%%%%%%%%%%%%%%%%%%%%%%%%%%%%%%%%%%%%%%%%%%%%%%%%%%%%%%%%%
%%%%%%%%%%%%%%%%%%%%%%%%%%%%%%%%%%%%%%%%%%%%%%%%%%%%%%%%%%%%%%%
%%%%%%%%%%%%%%%%%%%%%%%%%%%%%%%%%%%%%%%%%%%%%%%%%%%%%%%%%%%%%%%

%\bibliography{references.bib}% Produces the bibliography via BibTeX.

\begin{thebibliography}{53}
\bibitem [{not()}]{note2}%
  \BibitemOpen
  \href@noop {} {}\bibinfo {note} {Given a plane defined by equation $ax + by +
  z + c=0$ and a point of coordinates $((x_1,y_1,z_1))$, we define the signed
  distance of the point from the plane as $\widetilde{d} \equiv (ax_1 + by_1 +
  z_1 + c)/\sqrt{a^2 + b^2 + 1}$. The usual non-negative definite distance is
  $d=\vert \tilde{d} \vert$.}\BibitemShut {Stop}%
\end{thebibliography}

%%%%%%%%%%%%%%%%%%%%%%%%%%%%%%%%%%%%%%%%%%%%%%%%%%%%%%%%%%%%%%%
%%%%%%%%%%%%%%%%%%%%%%%%%%%%%%%%%%%%%%%%%%%%%%%%%%%%%%%%%%%%%%%
%%%%%%%%%%%%%%%%%%%%%%%%%%%%%%%%%%%%%%%%%%%%%%%%%%%%%%%%%%%%%%%

\end{document}